\documentstyle[11pt]{article}

\makeatletter
\def\dif{{\rm d}}
\def\deriv{\@ifnextchar[{\@deriv}{\@deriv[]}}
   \def\@deriv[#1]#2#3{\mathchoice%
{{\dif^{#1}#2\over\dif{#3}^{#1}}}{{\dif^{#1}#2/\dif{#3}^{#1}}}%
{{\dif^{#1}#2\over\dif{#3}^{#1}}}{{\dif^{#1}#2/\dif{#3}^{#1}}}}

\def\secteqno{\@addtoreset{equation}{section}%
\def\theequation{\thesection.\arabic{equation}}}
\newcounter{subequation}
\def\thesubequation{\alph{subequation}}
\def\sneqnarray{\stepcounter{equation}\let\@currentlabel=\theequation
\setcounter{subequation}{1}
\def\@eqnnum{{\rm (\theequation.\thesubequation)}}
\global\@eqcnt\z@\tabskip\@centering\let\\=\@eqncr\let\@@eqncr=\@@sneqncr
$$\halign to \displaywidth\bgroup\@eqnsel\hskip\@centering
 $\displaystyle\tabskip\z@{##}$&\global\@eqcnt\@ne
 \hskip 2\arraycolsep \hfil${##}$\hfil
 &\global\@eqcnt\tw@ \hskip 2\arraycolsep $\displaystyle\tabskip\z@{##}$\hfil
  \tabskip\@centering&\llap{##}\tabskip\z@\cr}
\def\endsneqnarray{\@@sneqncr\egroup $$\global\@ignoretrue}
\def\@@sneqncr{\let\@tempa\relax
   \ifcase\@eqcnt \def\@tempa{& & &}\or \def\@tempa{& &}
   \else \def\@tempa{&}\fi
     \@tempa \if@eqnsw\@eqnnum\stepcounter{subequation}\fi
     \global\@eqnswtrue\global\@eqcnt\z@\cr}
\def\nobiblabels{\def\@lbibitem[##1]##2{\@bibitem{##2}}}
\makeatother
\def\tabaddress#1{{\small\it\begin{tabular}[t]{c}#1\\[1.2ex]\end{tabular}}}

\def\ben{\begin{enumerate}}
\def\een{\end{enumerate}}
\def\beq{\begin{equation}}
\def\eeq{\end{equation}}
\def\bea{\begin{eqnarray}}
\def\eea{\end{eqnarray}}
\def\beann{\begin{eqnarray*}}
\def\eeann{\end{eqnarray*}}
\def\beasn{\begin{sneqnarray}}
\def\eeasn{\end{sneqnarray}}

\newtheorem{prop}{Proposition}

\def\UBECM{Departament d'Estructura i Constituents de la Mat\`eria\\
   Universitat de Barcelona\\
   Av.~Diagonal 647\\
   08028 Barcelona\\
   Catalonia, Spain}

\let\oldGamma=\Gamma
\def\Gamma{{\bf\oldGamma}}

\def\sub#1{\mathrel{\mathop{=}\limits_{#1}}}

\def\d{\delta}

\def\subs#1{\mathrel{\mathop{\simeq}\limits_{#1}}}
\title{Boundary conditions from boundary terms, Noether charges and the
trace K lagrangian in general relativity. \\
}

\author{Josep M. Pons$^a$
\\
\tabaddress{$^a$\UBECM}
\\[2mm]
\small e-mail: 
{\tt pons@ecm.ub.es}
}

\date{\today}

\begin{document}

\maketitle

\begin{abstract}
We present the Lagrangian whose corresponding action is the trace K action
for General Relativity. Although this Lagrangian is second order in the 
derivatives, it has no second order time derivatives and its 
behavior at space infinity in the asymptotically flat case is identical to
other alternative Lagrangians for General Relativity, like the gamma-gamma 
Lagrangian used by Einstein. We develop 
some elements of the variational principle for field theories with boundaries,
and apply them to second order Lagrangians, where we stablish the conditions 
---proposition 1--- for the conservation of the Noether charges. From this
general approach a pre-symplectic form is naturally obtained that 
features two terms, one from the bulk and another from the boundary. 
When applied to the trace K Lagrangian, we recover a pre-symplectic form 
first introduced using a different approach. We prove that 
all diffeomorphisms satisfying certain restrictions
at the boundary ---that keep room for a realization of 
the Poincar\'e group--- 
will yield Noether conserved charges. In particular, the computation of 
the total energy gives, in the asymptotically flat case, the ADM result. 
\end{abstract}

\section{Introduction}

Either way, boundary conditions {\sl from} boundary terms or boundary terms 
{\sl for} boundary conditions; 
boundary terms for the action and boundary conditions for the field
configurations go together. Whereas from a purely logical point of view
the boundary conditions emerge as a consequence of the boundary terms,
from the technical side the situation is often the opposite: 
one must look for the boundary terms that are necessary to
implement the desired ---or acceptable--- boundary conditions. 

This discussion applies directly to General Relativity (GR). 
It is well known that the boundary conditions required for the correct 
application of the variational principle in GR depend
upon the boundary terms exhibited by the action. Since divergence terms 
do not alter Einstein equations of motion, there is some freedom to write down
Lagrangians for GR that differ in some divergence terms from the original 
Einstein-Hilbert proposal. These divergence terms lead to different 
boundary terms for the action. 
To be specific, consider the original Einstein-Hilbert 
action (integration is in a 4-volume ${}^4\!V$ of spacetime)
\beq
S_{EH} = \int_{{}^4\!V} d^4\!x  \ \sqrt{|g|} R  \ . 
\label{ehaction}
\eeq
In this case, the boundary conditions for the application of the variational 
principle involve the fixation of some derivatives of the metric at the 
boundary. Instead, subtraction of a divergence from (\ref{ehaction})
allows to write the ``gamma-gamma'' action, first used by Einstein,
\beq
S_{\Gamma} = \int_{{}^4\!V} d^4\!x  \ \sqrt{|g|} g^{\mu\nu}
(\Gamma^\rho_{\mu\sigma} 
\Gamma^\sigma_{\rho\nu} - \Gamma^\rho_{\mu\nu} 
\Gamma^\sigma_{\rho\sigma})  \ ,
\label{gammaaction}
\eeq
which is first order in the spacetime derivatives and that has the possible 
advantage that the variational principle only requires 
to fix the metric at the boundary.

In field theory, the association --by means of the Noether theorem-- of 
symmetries at the level 
of the variational principle with conserved currents makes
the role of boundary terms (leading to boundary conditions for the fields),
either in the Lagrangian or the Hamiltonian formulation,
essential for the conservation of charges, as shown in the pioneering
work of \cite{Regge:1974zd} for GR (see also recent work in 
\cite{Katz:1997si,Julia:1998ys,Julia:2000er}). 
The role of boundary terms 
has also been stressed in recent years by the introduction of the concept 
of quasilocal charges in GR (see 
\cite{Brown:1993br} for general references, 
see also \cite{Chen:1999aw,Chang:1999wj,brown00}).

Many actions, all differing from Einstein-Hilbert's in 
boundary terms, have been used in the literature. 
With no aim of completeness, 
let us mention contributions in \cite{dirac58, gibb77,
york72,Charap:1983kn,york86,gold87,hayw93,Hawk96fd,hawk96,epp98,Franca02a,
Franca02b} 
and many references therein. Among them, and using the terminology of 
\cite{brown00}, the ``trace K'' action  
\cite{york72,york86,brown00} offers some singular features, because of 
its geometric 
character, that deserve full interest. This action is roughly written as
\beq
S_{K} = \int_{{}^4\!V} d^4\!x  \ \sqrt{|g|} R 
- 2\int_{\partial{}^4\!V} d^3\!x  \  \sqrt{|\gamma|} K \ , 
\label{trace}
\eeq
where $\partial{}^4\!V$ is the boundary of the 4-volume ${}^4\!V$, 
$\gamma$ is
the determinant of the 3-metric induced on the boundary and $K$ is its
extrinsic curvature. When the boundary is not smooth, and 
discontinuities arise in the vector field orthonormal to the boundary, 
the second term in (\ref{trace}) must be understood \cite{hayw93} 
as including delta-like 
contributions from the ``joints'' between the smooth elements of the
boundary. The detailed form of $S_{K}$ is given in section 3. 

The purpose of the present paper is to analyse the Lagrangian leading
to the action (\ref{trace}) and to provide with the theoretical framework
for actions of this type. In section 2 we introduce some notation and
useful formulas. The Lagrangian for the trace K action is obtained
in section 3. In section 4 we develop the formal theory for the 
variational principle and the Noether symmetries for second order 
Lagrangians in field theories with boundaries. This theory is applied
in section 5 to the trace K Lagrangian and we compute the total energy 
in the particular case of asymptotically flat spaces. Conclusions are
presented in section 6 and the appendix is devoted 
to prove some expressions used in the text.

\section{Some notation and formulas}

Here we introduce some notation and formulas to be used in the next 
sections. We consider that our spacetime coordinates 
correspond to a standard $3+1$ decomposition, including in particular that
 $g^{00} < 0$ 
and $g^{ii} > 0$, that is,  no surface $x^\mu = constant$ is tangent to the 
light cone.
Working with this type of coordinates means no restriction,
at least infinitesimally, on the gauge freedom, because any infinitesimal 
diffeomorphism preserves these conditions for the metric components. 
The determinant of the 4-metric $g_{\mu\nu}$ will be denoted by $g$.

Given the 3-surface $x^\mu = constant$, its orthonormal vector ${\bf n}(\mu)$
is defined by the relation
$$
n^\mu(\mu) n^\nu(\mu) = \xi(\mu) g^{\mu\nu} \ ,
$$ 
and by requiring that it is pointing towards increasing values of 
the coordinate $x^\mu$. 
The coefficient $\xi(\mu)$ is just a sign, $\xi(\mu) = \eta_{\mu\mu}$, 
where $\eta_{\mu\nu}$ is the Minkowski metric
with positive signature $(-,+,+,+)$. Then,
$$
n^\nu(\mu) = \xi(\mu) {g^{\mu\nu} \over \sqrt{|g^{\mu\mu}|}}
$$

Let us point out that ${\bf n}(\mu)$ is not a true vector field for it fails
to transform as such under the diffeomorphisms that do not preserve the 
foliation defined by the 3-surfaces $x^\mu = constant$. We compute in the 
appendix the deviation from the vector behavior of the transformation 
of ${\bf n}(\mu)$ under diffeomorphisms. 

The 3-metric induced at the 3-surface $x^\mu = constant$ is $g_{ab}$, with
$a,b = 0,1,2,3 \ except \  \mu$. Its inverse matrix is given by 
$\gamma^{ab}(\mu)$; the components of this matrix are the non identically 
vanishing components of
\beq
\gamma^{\rho\sigma}(\mu):=g^{\rho\sigma}-
{g^{\rho\mu}g^{\mu\sigma} \over g^{\mu\mu}} = 
g^{\rho\sigma}-\xi(\mu) n^\rho(\mu) n^\sigma(\mu).      
\label{gammag}
\eeq
The determinant $\det{g_{ab}}$ (for $a,b = 0,1,2,3 \ except \  \mu$) 
will be denoted $\gamma(\mu)$. There is the relationship 
$$
g g^{\mu\mu} = \gamma(\mu).
$$

Consider, for $\nu \neq \mu$, the 2-surface 
$x^\mu = constant, \ x^\nu = constant$. The induced metric on it is $g_{AB}$, 
with $A,B = 0,1,2,3 \ except \ \mu, \nu$. Its inverse metric will be written 
as $\gamma^{AB}(\mu\nu)$, and the determinant 
$\det{g_{AB}} =: \gamma(\mu\nu)$. One can show:
$$
g (g^{\mu\mu}g^{\nu\nu} - (g^{\mu\nu})^2 ) = 
g g^{\mu\mu} \gamma^{\nu\nu}(\mu) = \gamma(\mu\nu).
$$
  
Note also that 
\beq
\gamma^{\nu\nu}(\mu) = g^{\nu\nu} 
( 1 - {(g^{\mu\nu})^2 \over g^{\mu\mu}g^{\nu\nu}}) =
g^{\nu\nu}( 1 - \xi(\mu)\xi(\nu) q^2(\mu\nu)),
\label{gamm}
\eeq
where we have defined the scalar products
$$
q(\mu\nu) = { g^{\mu\nu} \over \sqrt{\xi(\mu)\xi(\nu)g^{\mu\mu}g^{\nu\nu}}}
= \xi(\mu)\xi(\nu) {\bf n}(\mu)\cdot{\bf n}(\nu) \ .
$$

The extrinsic curvature for the surface $x^\mu = constant$ is given by
$$
K_{ab}(\mu) := - {1\over \sqrt{|g^{\mu\mu}|}}\Gamma^\mu_{ab}   \ , 
$$
which is equivalent to
$$
K_{ab}(\mu)= {1\over 2} \xi(\mu) {\cal L}_{{\bf n}(\mu)} g_{ab} \ ,
$$
where ${\cal L}_{{\bf n}(\mu)}$ is the Lie derivative under ${\bf n}(\mu)$. 

The trace of the extrinsic curvature $K(\mu) := \gamma^{ab}(\mu) K_{ab}(\mu)$
may be written as
$$
K(\mu) = \xi(\mu) {\cal L}_{{\bf n}(\mu)}(ln \sqrt{|g|}) = 
\xi(\mu) n^\nu_{;\nu}(\mu) \ ,
$$
where $n^\nu_{;\nu}(\mu)$ is the covariant derivative of ${\bf n}(\mu)$,
$\nabla_\nu n^\nu(\mu)$, with the Riemannian connexion, 
as if ${\bf n}(\mu)$ were a true vector.
\section{The Lagrangian for the trace K action}

In this section we use techniques 
introduced in \cite{kijowski97} (see also\cite{kijowski84,kijowski86}). 
Let us first take a look on the boundary 
conditions imposed by the variational principle for the Einstein-Hilbert 
Lagrangian. A general variation for ${\cal L}_{EH}$ gives
\beq
\d {\cal L}_{EH} = - G^{\mu\nu} \d g_{\mu\nu} + 
\sqrt{|g|} g^{\mu\nu} \d R_{\mu\nu},
\label{vareh} 
\eeq   
where $R_{\mu\nu}$ stands for the Ricci tensor and 
$$
G^{\mu\nu} := \sqrt{|g|} (R^{\mu\nu} - {1\over 2}g^{\mu\nu} R), 
$$  
with $R$ the scalar curvature.

As it is well known, the last term in (\ref{vareh}) is a divergence. 
It can be written as
\beq
\sqrt{|g|} g^{\mu\nu} \d R_{\mu\nu} = 
\partial_\mu(\tilde g^{\rho\nu\mu}_\sigma \d \Gamma^\sigma_{\rho\nu}) \ ,
\label{thediverg}
\eeq
where $\tilde g^{\rho\nu\mu}_\sigma:=
\sqrt{|g|} (g^{\rho\nu} \d^\mu_\sigma - g^{\mu (\nu}\d^{\rho)}_\sigma).
$
The next step is to write \cite{kijowski97} 
$\tilde g^{\rho\nu\mu}_\sigma \d \Gamma^\sigma_{\rho\nu}$ as 
(we continue with the convention of indices $a,b = 0,1,2,3 \ except \  \mu$)
\beq
\tilde g^{\rho\nu\mu}_\sigma \d \Gamma^\sigma_{\rho\nu} =
- g_{ab} \d P^{ab}(\mu) + \partial_\nu (\sqrt{|g|} g^{\mu\mu} 
\d ({g^{\mu\nu}\over g^{\mu\mu}})) \ ,
\label{kijfirst}
\eeq
where $P_{ab}(\mu)$ (for a given $\mu$, notice that 
we raise and lower indices with 
$g_{ab}, \gamma^{ab}(\mu)$)
is a generalisation to any $\mu$ of the ADM \cite{arnowitt/deser/misner/62}
 momenta for $\mu=0$:
$$
P_{ab}(\mu) = \sqrt{|\gamma(\mu)|}(g_{ab} K(\mu) - K_{ab}(\mu)).
$$
Therefore, the divergence term in (\ref{thediverg}) is 
\bea
\partial_\mu(\tilde g^{\rho\nu\mu}_\sigma \d \Gamma^\sigma_{\rho\nu}) 
&=& - \partial_\mu(g_{ab}(\mu) \d P^{ab}(\mu)) + 
\partial_\mu \partial_\nu \left(\sqrt{|g|} g^{\mu\mu} 
\d ({g^{\mu\nu}\over g^{\mu\mu}})\right)  \nonumber \\
&=& - \partial_\mu(g_{ab}(\mu) \d P^{ab}(\mu)) + 
{1\over 2} \partial_\mu \partial_\nu \big(\sqrt{|g|} \left( g^{\mu\mu} 
\d ({g^{\mu\nu}\over g^{\mu\mu}}) + g^{\nu\nu} 
\d ({g^{\mu\nu}\over g^{\nu\nu}})\right)\big). 
\eea

Finally, the term acted upon by the derivatives $\partial_\mu \partial_\nu$ 
has the equivalent expression:
\bea
{1\over 2} \sqrt{|g|} \left( g^{\mu\mu} 
\d ({g^{\mu\nu}\over g^{\mu\mu}}) + g^{\nu\nu} 
\d ({g^{\mu\nu}\over g^{\nu\nu}})\right) &=& 
\left(\sqrt{|g|} \sqrt{\xi(\mu)\xi(\nu) g^{\mu\mu}g^{\nu\nu}}\right)
\d \left( {g^{\mu\nu}\over \sqrt{\xi(\mu)\xi(\nu) g^{\mu\mu}g^{\nu\nu}}} 
\right) \nonumber \\ 
&=& 
\left(\sqrt{|g|} \sqrt{\xi(\mu)\xi(\nu) g^{\mu\mu}g^{\nu\nu}}\right)
\d q(\mu\nu)  \nonumber \\ 
&=&
\sqrt{|\gamma(\mu\nu)|} \d \alpha(\mu\nu) \ ,
\label{angles}
\eea
where the angles $\alpha(\mu\nu)$ are defined as
$$
\alpha(0i) = {\rm arcsinh} (q(0i)) \ ,  
\quad \alpha(ij) = {\rm arcsin} (q(ij)) \ ,
$$ 
and we have used  
$$
\d \alpha(\mu\nu) = {\d q(\mu\nu) \over 
\sqrt{( 1 - \xi(\mu)\xi(\nu) q^2(\mu\nu))}} \ .
$$
Therefore
\beq
\d S_{EH} = \int_{{}^4\!V} d^4\!x  \ \d {\cal L}_{EH} 
= \int_{{}^4\!V} d^4\!x \  \left( - G^{\mu\nu} \d g_{\mu\nu} 
- \partial_\mu(g_{ab}(\mu) \d P^{ab}(\mu)) + \partial_\mu \partial_\nu
(\sqrt{|\gamma(\mu\nu)|} \d \alpha(\mu\nu)) \right)
\label{deleh}
\eeq
Consider the volume ${}^4\!V$ being a 4-cube, having as boundary elements 
3-surfaces
defined by the constancy of one of the coordinates $x^\mu$. Then the second
term in the integrand will give a contribution from the faces and the third
a contribution from the joints, these joints being defined by the constancy 
of two coordinates.
In order for $S_{EH}$ to be a differentiable functional \cite{brownhenn86}, 
and its 
variation to consequently lead to Einstein equations, we need to control the 
fields and the variations at the boundary in such a way that
$$
 \d S_{EH} = 0 \Longleftrightarrow  G^{\mu\nu} =0.
$$ 

As we have already said, this control at the boundary involves some derivatives
of the metric components. Now we have the specifics: the variation of 
$P^{ab}(\mu)$ must vanish at the faces $x^\mu = constant$, and the 
variation of $\alpha(\mu\nu)$ must vanish at the joints between the
faces $x^\mu = constant$ and $x^\nu = constant$. The physical meaning
of this restrictions on the variations is unclear.

Instead, a simple addition of a divergence term to ${\cal L}_{EH}$ will give
a more reasonable control of the variations at the boundary. Taking into 
account that $g_{ab}(\mu)  P^{ab}(\mu) = 2 \sqrt{|\gamma(\mu)|} K(\mu)$, 
define the new Lagrangian ($K$ is for the extrinsic curvature)
\beq
{\cal L}_{K} := {\cal L}_{EH} + 2 \partial_\mu(\sqrt{|\gamma(\mu)|} K(\mu))
-  \partial_\mu \partial_\nu(\sqrt{|\gamma(\mu\nu)|}  \alpha(\mu\nu)) \ .
\label{tracelag}
\eeq
The action (\ref{trace}) is, by definition,  
$S_{K} := \int_{{}^4\!V} d^4\!x  \  {\cal L}_{K}$. Then,
\beq 
\d S_{K} = \int_{{}^4\!V} d^4\!x  \ \d {\cal L}_{K}   
= \int_{{}^4\!V} d^4\!x  \ \left( - G^{\mu\nu} \d g_{\mu\nu} 
+ \partial_\mu(P^{ab}(\mu) \d g_{ab}(\mu) ) - \partial_\mu \partial_\nu
(\alpha(\mu\nu) \d \sqrt{|\gamma(\mu\nu)|} ) \right)
\label{vartracelag}
\eeq

So for $S_{K}$ to be a differentiable functional we must require, 
besides the customary vanishing of the variations of the 3-metric induced on 
the initial and final 
equal-time 3-surfaces, the supplementary condition of the vanishing
of the variations of the 3-metric induced on any spatial 3-face of the 
boundary. These requirements already 
guarantee the fixation of the variations of the determinant $\gamma(\mu\nu)$ 
at the joints. In fact, to be more precise, the vanishing of the variations
of the 3-metric is only required for finite boundaries; in the case
of asymptotically flat spaces, where we let the spatial 
elements of the boundary go to the space infinity ($r \rightarrow \infty$), 
we must control
the $r \rightarrow \infty$ behavior of the fields and its 
allowed variations 
so that there is no contribution to $\d S_{K}$. More on this later.

Notice the difference with the boundary conditions that one finds in the case
of the gamma-gamma Lagrangian, where the vanishing of the variations
of the 4-metric is required. 

Expression (\ref{tracelag}) is the Lagrangian for the action (\ref{trace}). 
We notice that the contributions from the joints are all included. The role 
of these contributions has been explained in \cite{hayw93}. Early 
computations can be found in  \cite{kijowski84,kijowski86}.

Two features of (\ref{tracelag}) are worth being mentioned immediately. 
First, as it happens generally with the proposals to modify 
${\cal L}_{EH}$ through
divergence terms, ${\cal L}_{K}$ is not a truly scalar density. 
This fact does not opposes to its physical applicability, as it is argued in
a parallel context in \cite{faddeev82} for the gamma-gamma Lagrangian.  
${\cal L}_{K}$ gives indeed the trace 
K action, with all its geometric meaning, but for coordinates 
adapted to the
boundary --or viceversa: for boundaries adapted to the coordinates--, 
in such a way that the elements of the boundary correspond to the
constancy of the value of some coordinate. A typical boundary may be the 
4-cube, already used, but a cylinder whose top and bottom faces are 
equal-time 3-surfaces, and whose lateral face is defined by the constancy
of a single radial coordinate, is a well adapted boundary as well. In the 
cylinder case the $(ij)$ ($i,j$ is for space indices) 
contributions from the last term in 
(\ref{vartracelag}) disappear in $S_K$ because ``the boundary of a 
boundary is zero''. 

The second observation is that,
unlike the ``gamma-gamma'' Lagrangian (\ref{gammaaction}), 
${\cal L}_{K}$ is not a first order Lagrangian. This is a point not 
sufficiently recognised in the literature 
\cite{gibb77,york86,brill92,Marolf:1995cp}. 
It is proved in the appendix that the second 
order contributions to ${\cal L}_{K}$ are as follows:
\beq
{\cal L}_{K} = ({\rm quadratic \ \, terms \ \, in \ \, the \ \, first \ \, 
derivatives \ \,  of \ \, the \ \, metric})
- \alpha(\mu\nu) 
\partial_\mu \partial_\nu\sqrt{|\gamma(\mu\nu)|},
\label{secorder}
\eeq
but note a key difference with the Einstein-Hilbert Lagrangian: 
${\cal L}_{K}$  has no second order time derivatives (the sum $\mu,\nu$ in
(\ref{secorder}) only contributes for $\mu\neq\nu$). In this sense 
${\cal L}_{K}$ has an intermediate place between 
${\cal L}_{EH}$ (with second order time derivatives) and 
${\cal L}_{\Gamma}$ (with only first order spacetime derivatives). 
In the next section we will introduce some notation and results for theories 
with second order Lagrangians to focus later on Lagrangians of the type of
${\cal L}_{K}$. The advantage of not having
second order time derivatives will become clear when we analyse the boundary 
conditions. Also, we will see in the asymptotically flat case that the
long distance behavior of ${\cal L}_{K}$ improves crucially that of
${\cal L}_{EH}$.  

\section{Second order Lagrangians: variational principle for field 
theories with boundaries}

Here we will consider a generic second order Lagrangian density function 
${\cal L}$ with dependences:
$$
{\cal L}(\phi, \phi_\mu, \phi_{\mu\nu})
$$
$\phi$ denotes the whole set of fields (a new index could
be introduced but it will be unnecessary). $\phi_\mu$ stands for 
$\partial_\mu\phi$, and $\phi_{\mu\nu}:= \partial_\mu\partial_\nu\phi$. 
If, as it happens with ${\cal L}_{K}$, only terms with $\mu\neq\nu$ appear 
in the second derivatives, some of the expressions simplify somewhat.

The Lagrangian functional is  
$$
L[\phi, \dot \phi,\ddot \phi] = \int_{{}^3\!V} d^3\!x  \ 
{\cal L}(\phi, \phi_\mu, \phi_{\mu\nu}),
$$
for some spatial 3-volume, 
and the action functional is
$$
S[\phi] =  \int_{t_0}^{t_1} dt   \  L[\phi, \dot \phi,\ddot \phi ] = 
\int_{{}^4\!V} d^4\!x \  {\cal L}(\phi, \phi_\mu, \phi_{\mu\nu}),
$$
with ${{}^4\!V}=[t_0, t_1] \times {{}^3\!V}$.
The functional differentiation of a general S is given \cite{soloviev} by
\beq
\d S = \int d^4\!x  \ \partial^{(m)}\left({\d S \over \d \phi^{(m)}} 
\d \phi \right)
\label{dels}
\eeq
($m$ is a condensed notation for any number of partial derivatives with 
respect to any spacetime coordinate),
that must be compared with 
\beq
\d S = \int d^4\!x  \ \d {\cal L} = \int d^4\!x  \ 
{\partial {\cal L} \over \partial \phi^{(m)}} \d \phi^{(m)}
\label{dell}
\eeq
in order to define the functional derivatives ${\d S \over \d \phi^{(m)}}$.
We will ignore, in the way of computing variations with respect to the 
second order derivatives, the symmetry
$\phi_{\mu\nu} = \phi_{\nu\mu}$. This means that we compute independently, for
instance, $\d S \over \d \phi_{\mu\nu}$ and  $\d S \over \d \phi_{\nu\mu}$. It 
is useful then to define the ``complete'' derivative ---that takes into account
the truly independent variables---,
$$
{\d^c S \over \d \phi_{\mu\nu}} = {\d^c S \over \d \phi_{\nu\mu}} := 
{\d S \over \d \phi_{\mu\nu}}+ {\d S \over \d \phi_{\nu\mu}},
$$
for $\mu\neq\nu$, and 
$$
{\d^c S \over \d \phi_{\mu\mu}} = {\d S \over \d \phi_{\mu\mu}}
$$
for the rest.

Comparing (\ref{dels}) and (\ref{dell}) gives, 
\bea
{\d S \over \d \phi} &=& {\partial {\cal L} \over \partial \phi}
- \partial_\mu ({\partial {\cal L} \over \partial \phi_\mu})
+ \partial_\nu \partial_\mu({\partial {\cal L} \over \partial \phi_{\mu\nu}}) 
\\
{\d S \over \d \phi_\mu}&=& {\partial {\cal L} \over \partial \phi_\mu}
- \partial_\nu({\partial {\cal L} \over \partial \phi_{\mu\nu}}
+{\partial {\cal L} \over \partial \phi_{\nu\mu}}) \\
{\d S \over \d \phi_{\mu\nu}} &=& 
{\partial {\cal L} \over \partial \phi_{\mu\nu}} \ .
\eea
The first line displays the Euler-Lagrange derivatives,   
yielding the equations of motion when set to zero. 

Analogous variations can be computed for $L[\phi, \dot \phi, \ddot \phi]$, 
taking into 
account that $\phi$, $\dot \phi$ and $\ddot \phi$ are independent 
arguments for $L$. It is convenient to define
$$
{\partial {\cal L} \over \partial \dot \phi_{i}} :=
{\partial {\cal L} \over \partial \phi_{0i}}+
{\partial {\cal L} \over \partial \phi_{i0}} , \ \ \
{\d S \over \d \dot \phi_{i}} :=
{\d^c S \over \d \phi_{0i}}, \ \ \ {\d L \over \d \dot \phi_{i}} :=
{\d^c L \over \d \phi_{0i}} \ .
$$
For the sake of completeness we give the relationship between functional
derivatives for $S$ and $L$. They are 
$$
{\d S \over \d  \phi} = {\d L \over \d  \phi} - 
\partial_0({\d L \over \d  \dot \phi}) + 
\partial_0\partial_0({\d L \over \d  \ddot \phi}), \ \ \ 
{\d S \over \d  \dot \phi} = {\d L \over \d  \dot\phi} 
- 2 \partial_0({\d L \over \d  \ddot \phi}), \ \ \
{\d S \over \d  \phi_i} ={\d L \over \d  \phi_i} - 
\partial_0({\d L \over \d  \dot \phi_i}),
$$
and
$$
{\d S \over \d \dot \phi_{i}} = {\d L \over \d \dot \phi_{i}}, \ \ \
{\d S \over \d \phi_{ij}} = {\d L \over \d \phi_{ij}} \ .
$$

Now we will explore the requirement of differentiability coming from
the variational principle. Using (\ref{dell})
\bea
\d S &=& \int_{{}^4V} d^4\!x  \ \left({\d S \over \d  \phi}  \d  \phi
+ \partial_\mu( {\d S \over \d  \phi_\mu} \d  \phi)
+ \partial_\nu\partial_\mu ({\d S \over \d  \phi_{\mu\nu}}\d  \phi)\right)  
\nonumber \\
&=& \int_{{}^4V} d^4\!x  \ {\d S \over \d  \phi} \d  \phi
+ \left[\int_{{}^3V} d^3\!x  \ {\d S \over \d \dot \phi} \d \phi 
\right]^{t=t_1}_{t=t_0}
+ \int_{t_0}^{t_1} dt \int_{\partial{}^3V}  d\sigma_i  \ 
{\d S \over \d \phi_{i}}\d  \phi  
\nonumber \\
&+& \left[\int_{\partial{}^3V} d\sigma_i  \ 
{\d S \over \d \dot \phi_{i}}\d  \phi \right]^{t=t_1}_{t=t_0}
+ \int_{{}^4V} d^4\!x  \ \partial_i\partial_j({\d S \over \d \phi_{ij}} 
\d  \phi) + \left[\int_{{}^3V} d^3\!x  \ 
\partial_0({\d S \over \d  \phi_{00}} \d \phi) 
\right]^{t=t_1}_{t=t_0} \ ,
\label{delsabstract}
\eea
where ${\partial{}^3V}$ is the --spatial-- boundary of ${{}^3V}$ and 
$d\sigma_i$ are the two-forms induced at the boundary by the application of 
Stokes theorem.

The second term, the fourth term, and a piece within
the sixth term, in the last equality, vanish because 
$\d \phi|_{t_0} = \d \phi|_{t_1} = 0$ as conditions imposed on the variations
allowed by the variational principle. So in order for the variation of $S$
to determine the Euler-Lagrange equations, 
$$
\d S = 0 \longleftrightarrow {\d S \over \d  \phi} = 0 \ ,
$$
we need 
\beq
\int_{t_0}^{t_1} dt \int_{\partial{}^3V} d\sigma_i \  
\left({\d S \over \d \phi_{i}}\d  \phi 
+ \partial_j({\d S \over \d \phi_{ij}} \d  \phi)\right) 
+ \left[\int_{{}^3V} d^3\!x  \ 
({\d S \over \d  \phi_{00}} \d \dot \phi) 
\right]^{t=t_1}_{t=t_0}
= 
0.
\label{boundco}
\eeq

Note in the last term the presence of $\d \dot \phi$. This term complicates
the setting of the boundary conditions for the fields. Formula (\ref{boundco})
---or better, formula (\ref{delsabstract}), before the cancellations 
originated from the variational principle--- is responsible for the 
expression (\ref{deleh}) obtained for the Einstein-Hilbert action. 
Things become a lot easier if we assume our Lagrangian not to contain 
second order time derivatives, which is the case for ${\cal L}_K$. 
We will continue with this assumption.

\subsection{Second order Lagrangians with no second order time derivatives}

This implies that now $L$ is $L[\phi, \dot \phi]$ only. 
In this case, (\ref{boundco}) simplifies to
\beq
\int_{t_0}^{t_1} dt \int_{\partial{}^3V} d\sigma_i \  
\left({\d S \over \d \phi_{i}}\d  \phi 
+ \partial_j({\d S \over \d \phi_{ij}} \d  \phi)\right) 
= 
0,
\label{boundco2}
\eeq
and since the ---finite--- range for the time integration is arbitrary, we 
end up with
\beq
\int_{\partial{}^3V} d\sigma_i \  
\left({\d S \over \d \phi_{i}}\d  \phi 
+ \partial_j({\d S \over \d \phi_{ij}} \d  \phi)\right) 
= 
0.
\label{boundcond}
\eeq

Condition (\ref{boundcond}) must be fulfilled, {\sl at any time},  
by the fields and its variations in order to comply with the 
variational principle. Two restrictions originate from (\ref{boundcond}):
\begin{itemize}
\item
A restriction on the space of field configurations 
---spatial boundary conditions for the fields.
\item
Consequently, a restriction on the allowed variations so 
that the space of field configurations is preserved under such variations. 
\end{itemize}
Both restrictions altogether must imply (\ref{boundcond}). Obviously the
simplest restrictions one can think of are to fix the values of
the fields at the spatial boundary so that its variations at the 
boundary vanish (maybe not all the fields need to be fixed at the 
spatial boundary, because some of the functional derivatives of $S$ in 
(\ref{boundcond}) may vanish identically).
This trivially complies with (\ref{boundcond}). 

\vspace{4mm}

The restrictions set on the space of field configurations are subject 
to a consistency test: they must be compatible with the equations of motion,
\beq
{\d S \over \d  \phi} =0.
\label{thedyn}
\eeq
This compatibility is generally nontrivial, and may cause the rejection
of some boundary conditions when they do not agree with the dynamics
(\ref{thedyn}) \footnote{We leave aside the case when the boundary 
is an interface between two different physical regimes.}. 
To implement the compatibility of (\ref{boundcond}) with
(\ref{thedyn}), that must hold at any time, we can run, for a fixed time that
can be taken as the initial time ---when the initial conditions are set---, 
a Dirac-like algorithm of stabilisation of
constraints. Then we can end up, in principle, with secondary boundary 
conditions, tertiary, etc., that play the role of gauge fixing constraints. 
We refer to \cite{Pons:2000xt} for a form of the standard 
Dirac algorithm for obtaining of the secondary, 
tertiary, etc., constraints, that makes no mention to tangency conditions.
The idea that boundary conditions must be treated as Dirac constraints
is clear from the developments above and has been put forward in 
\cite{sheikh99}.
  
We have all things ready for the study of the implementation of the 
Noether symmetries for ${\cal L}_{K}$. 
Before finishing this section, let us mention that the first line of
equation (\ref{delsabstract}) ---or the more general version (\ref{dels})---, 
together with the variation of ${\cal L}_{K}$ in 
(\ref{vartracelag}), dictates the results:

\beq
{\d S_K \over \d  g_{ab,\mu}} = P^{ab}(\mu)
\label{bulkmom}
\eeq
for $a,b = 0,1,2,3  \ except \ \mu$, and
\beq
{\d S_K \over \d g_{AB,\mu\nu}} + {\d S_K \over \d g_{AB,\nu\mu}}
=  - \alpha(\mu\nu) \sqrt{|\gamma(\mu\nu)|}\gamma^{AB}(\mu\nu)
\label{boundmom}
\eeq
for $A,B = 0,1,2,3 \ except \ \mu, \nu$. 
Note that (\ref{boundmom}) is in agreement with (\ref{secorder}).
These results will be used in section 5.

\subsection{Noether currents and charges}

We first continue developing the general theory with the same assumptions 
of the preceding section: $L$ is $L[\phi, \dot \phi]$ and ${\cal L}$ is
${\cal L}(\phi, \dot \phi,\phi_i, \dot\phi_i, \phi_{ij})$

Now consider the Noether case. Suppose that our Lagrangian satisfies, under
a certain variation $\d_D$ 
(not necessarily complying with (\ref{boundcond})), 
\beq
\d_D {\cal L} =\partial_\mu F^\mu \ ,
\label{noethercase}
\eeq
for certain functions $F^\mu$. From the first line of (\ref{delsabstract}), 
$$
\d_D {\cal L} =  {\d S \over \d  \phi}  \d_D  \phi
+ \partial_\mu( {\d S \over \d  \phi_\mu} \d_D  \phi)
+ \partial_\nu\partial_\mu ({\d S \over \d  \phi_{\mu\nu}}\d_D  \phi), 
$$
we get
\beq
{\d S \over \d  \phi}  \d_D  \phi = \partial_\mu \left(
 F^\mu - {\d S \over \d  \phi_\mu} \d_D  \phi -
\partial_\nu({\d S \over \d  \phi_{\mu\nu}}\d_D  \phi) \right), 
\label{noether}
\eeq 
which identifies the current
\beq
J^\mu := F^\mu - {\d S \over \d  \phi_\mu} \d_D  \phi -
\partial_\nu({\d S \over \d  \phi_{\mu\nu}}\d_D  \phi) + 
\partial_\nu A^{\mu\nu},
\label{defcurr}
\eeq
as an on shell conserved object,  
\beq
\partial_\mu J^\mu  \sub{\rm (on \ shell)} 0 \ .
\label{conscurr}
\eeq

Note that we have included in $J^\mu$ the unavoidable ambiguity 
of the addition of the divergence of an arbitrary antisymmetric object
$A^{\mu\nu}$. 
When searching for a conserved 
charge, it will prove useful to take into account this ambiguity. 
Let us integrate (\ref{conscurr}) along the spatial 3-volume,
\beq
\partial_0 \int_{{}^3V} d^3\!x  \ J^0 
+ \int_{\partial{}^3V} d\sigma_i  \ J^i 
\sub{\rm (on \ shell)} 0,
\label{chargecurr}
\eeq
so the charge 
\beq
Q:=\int_{{}^3V} d^3\!x  \ J^0 
\label{thecharge}
\eeq
will be conserved on shell if 
$$
\int_{\partial{}^3V} d\sigma_i  \  J^i 
\sub{\rm (on \ shell)} 0.
$$
It is advantageous to take, in $J^\mu$, 
$$
A^{i0} = - A^{0i} = {\d S \over \d  \phi_{i0}}\d_D  \phi + B^{i0} \ , 
\quad  A^{ij} = B^{ij} \ , 
$$
with $B^{\mu\nu}$ another arbitrary antisymmetric object. 
Then the components of the current are , 
$$
J^0 = F^0 - {\d S \over \d \dot \phi} \d_D  \phi
- \partial_i({\d S \over \d \dot \phi_{i}}\d_D \phi - B^{0i})
$$
and
$$
J^i = F^i - 
{\d S \over \d  \phi_i} \d_D  \phi -
\partial_j({\d S \over \d  \phi_{ij}}\d_D  \phi)
+ \partial_\nu B^{i\nu} \ .
$$

Thus, 
$$
\int_{\partial{}^3V} d\sigma_i  \  J^i  =
\int_{\partial{}^3V} d\sigma_i  \ \left(F^i - 
{\d S \over \d  \phi_i} \d_D  \phi -
\partial_j({\d S \over \d  \phi_{ij}}\d_D  \phi)
+ \partial_\nu B^{i\nu}
)\right) \ ,
$$
and now (\ref{boundcond}) comes into play:  
the addition of the second and third terms in the right side vanish 
{\sl if the field configurations and $\d_D$ comply with (\ref{boundcond})}.
Here we realize the relevance of the boundary conditions when 
considering the conservation of charge. Form now on we assume that the 
variational principle has restricted 
our space of field configurations and the allowed 
variations $\d_D$ defined on it in such way that (\ref{boundcond}) 
is satisfied. 
Then the charge conservation (\ref{thecharge}) has the condition 
$$
\int_{\partial{}^3V} \  d\sigma_i (F^i  
+ \partial_\nu B^{i\nu}) )\sub{\rm (on \ shell)} 0
$$
whereas the charge itself takes the form
$$
Q:=\int_{{}^3V} d^3\!x  \ J^0  = 
\int_{{}^3V} d^3\!x  \ \left( F^0 - {\d S \over \d \dot \phi} \d_D  \phi
- \partial_i({\d S \over \d \dot \phi_{i}}\d_D \phi - B^{0i} ) \right).
$$
The presence of the object $B^{\mu\nu}$ may be endowed to the ambiguity 
of $F^\mu$ in (\ref{noethercase}), under additions of the divergence 
of an arbitrary antisymmetric object. 
Absorbing $\partial_\nu B^{i\nu}$ within $F^i$ and 
$\partial_i B^{0i}$ within $F^0$, we arrive at the 
following
\begin{prop}
Given a second order Lagrangian with no second order time derivatives. If
\begin{enumerate}
\item
the space of field configurations 
---including spatial boundary conditions for the fields---, 
together with the allowed 
variations defined on it, complies with (\ref{boundcond}),  
\item
the boundary conditions imposed on the field 
configurations are compatible with the equations of motion (\ref{thedyn}),
\item
there is an allowed variation $\d_D$ such that 
$\d_D {\cal L} =\partial_\mu F^\mu$ for some $F^\mu$, 
\end{enumerate}
then, the charge 
\beq
Q = \int_{{}^3V} d^3\!x \  ( F^0 - {\d S \over \d \dot \phi} \d_D  \phi)
- \int_{\partial{}^3V} d\sigma_i \  ({\d S \over \d \dot \phi_{i}}\d_D \phi)
\label{truecharge}
\eeq
is conserved if and only if the condition 
\beq
\int_{\partial{}^3V} d\sigma_i \  F^i  \sub{\rm (on \ shell)} 0 \ ,
\label{condcons}
\eeq
holds.
\end{prop}

For infinite boundaries, as in the case of the limit $r \rightarrow \infty$
for asymptotically flat spaces, (\ref{condcons}) is understood as
\beq
 \lim_{r \rightarrow \infty} \int_{\partial{}^3V} d\sigma_i \  F^i  
\sub{\rm (on \ shell)} 0 \ .
\label{condconsinf}
\eeq

Note the generic form of the charge ---conserved if the conditions of 
proposition 1 hold--- in (\ref{truecharge}): a bulk term and a boundary
term, the existence of this last term having its origin in the dependence 
of the Lagrangian on the second order derivatives.
 
It is worth to remark that the Noether current conservation (\ref{conscurr}) 
stemming from (\ref{noether}) is a local property {\sl only} of the equations 
of motion, and it is totally impervious to the modifications of the 
action by boundary terms. The Noether current is local, and for its
definition (\ref{defcurr}) it is completely irrelevant whether the
variation $\d_D$ producing (\ref{noethercase}) is an allowed variation 
---in the sense of (\ref{boundcond})--- or not.
Let the action be $S_{EH}$, $S_\Gamma$ or $S_K$ 
---each leading to different boundary conditions---, the current conservation 
equation (\ref{conscurr}) will always be the same (but remember: there is
an ambiguity in $J^\mu$, to wit, the addition of 
a divergence of an arbitrary antisymmetric object). What then is the 
relevance of using one action or another? the answer is that
the boundary terms in the action are indeed relevant to set the conditions 
(\ref{boundcond}). Selecting the action entails a selection of the space of 
field configurations and the allowed variations on it
---though maybe not in a unique way. 
It is obvious then, looking at the proposition above, that the charge 
conservation crucially depends on the selected action.

If the current conservation (\ref{conscurr}) holds but the conditions of 
Proposition 1 are not satisfied, then the equation (\ref{chargecurr}) will
express the typical conservation equation that balances the rate of change 
in time of the total charge with the flux of current through 
the boundary.

\subsection{Energy and (pre)symplectic structure}

Here we continue with a second order Lagrangian ${\cal L}$ with no 
second order time derivatives. 
Since there is no explicit dependence of ${\cal L}$ on the 
coordinates,
$\d_D \phi = \dot \phi \d t$, with $\d t$ an infinitesimal constant, 
is a Noether
symmetry with $F^\mu = \d^\mu_0 {\cal L}\d t$. Since $F^i = 0$, 
the conservation
of (\ref{truecharge}) will always hold {\sl as long as conditions 1) and 2)
in Proposition 1 are satisfied}. With this proviso,  
the conserved quantity (\ref{truecharge}) is (up to a sign) the energy:
\beq
E = \int_{{}^3V} d^3\!x  \ (  {\d S \over \d \dot \phi} \dot  \phi 
- {\cal L})
+ \int_{\partial{}^3V} d\sigma_i  \ 
({\d S \over \d \dot \phi_{i}}\dot \phi) \ ,
\label{energy}
\eeq
that can also be expressed in terms of the Lagrangian functional,
$$
E = \int_{{}^3V} d^3\!x  \ (  {\d L \over \d \dot \phi} \dot  \phi )
+ \int_{\partial{}^3V} d\sigma_i  \ ({\d L \over \d \dot \phi_{i}}\dot \phi)
- L[\phi, \ \dot \phi].
$$

Expression (\ref{energy}) is the formula for the energy for second order
Lagrangians with no second order time derivatives. It generalises
the common expression $E = \hat p_k \dot q^k - L(q, \dot q)$, corresponding
to the Legendre transformation in mechanics, where 
$$\hat p_k := {\partial L \over \partial \dot q^k}$$
is the pullback to tangent space of the momentum in cotangent 
space (phase space) through the Legendre map. Some interesting consequences 
may be drawn from this generalisation. Summation for an index $k$ in
mechanics becomes in field theory integration for the space coordinates 
plus summation for all 
the fields. Using this mechanical analogy, 
the role of the pullback $\hat p_k d q^k$
of the Liouville one-form is now played by 
$$\int_{{}^3V} d^3\!x  \ (  {\d S \over \d \dot \phi} \d  \phi) 
+ \int_{\partial{}^3V} d\sigma_i  \ ({\d S \over \d \dot \phi_{i}}\d \phi) \ ,
$$
which can indeed be called the pullback of the Liouville form for this 
type of field theory.
The pullback $ d \hat p_k \wedge d q^k$ to tangent space of the 
symplectic two-form in phase space now becomes
\beq
\hat \Omega = \int_{{}^3V} d^3\!x \  ( \d {\d S \over \d \dot \phi}
\wedge \d  \phi) 
+ \int_{\partial{}^3V} d\sigma_i \  
(\d {\d S \over \d \dot \phi_{i}}\wedge\d \phi) \ .
\label{presympl}
\eeq
This structure can be symplectic (closed and maximal rank) 
or presymplectic (non maximal rank) 
depending on the presence of gauge symmetries in the theory. 

Expression (\ref{presympl}) also means that each canonical momentum has 
now two components: the bulk component and the boundary component. Their
pullbacks to tangent space are, respectively, 
\beq
p_{({\rm bulk})} = {\d S \over \d \dot \phi}
\label{bulkm}
\eeq
and 
\beq
p^i_{({\rm boundary})}= {\d S \over \d \dot \phi_{i}} \ .
\label{boundm}
\eeq

\section{Application to ${\cal L_K}$: asymptotically flat spaces}

Here we follow Faddeev approach \cite{faddeev82}. Asymptotically flat 
spacetimes correspond  
to physical situations where the gravitating masses and matter fields
at finite times are effectively concentrated in a finite region of space.
Our spacetime will be a topologically simple manifold whose points 
can be parametrised by a system of four coordinates 
$x^\mu, \, -\infty < x^\mu < \infty $, 
such that, in the limit $r \rightarrow \infty $ 
($ r^2 = (x^1)^2+(x^2)^2+(x^3)^2$) for finite time $t = x^0$, the metric 
components satisfy the asymptotic conditions 
\bea
g_{\mu\nu} &=& \eta_{\mu\nu} + O({1 \over r}), \quad \,
\partial_\sigma g_{\mu\nu} = O({1 \over r^2}), \nonumber \\
\partial_{\sigma\rho} g_{\mu\nu} &=& O({1 \over r^{3}}),
\ \ \cdots ,  \partial^{(m)} g_{\mu\nu} = O({1 \over r^{(m+1)}}) .
\label{asym}
\eea  

Conditions (\ref{asym}) amount to a partial gauge fixing, for the only  
acceptable changes of coordinates will be from now on those that 
preserve (\ref{asym}). To this end, for an infinitesimal
change $x^\mu \rightarrow x'^\mu = x^\mu - \epsilon^\mu(x)$, we take 
$\epsilon^\mu$ to behave as
\bea
\epsilon^\mu &=& \omega^\mu_{\ \nu} x^\nu + a^\mu + O({1 \over r}), 
\nonumber \\
\partial_\nu\epsilon^\mu &=& \omega^\mu_{\ \nu} + O({1 \over r^2})
\nonumber \\
\partial_\nu \partial_\sigma\epsilon^\mu &=& O({1 \over r^{2 + \alpha}}),
\, \alpha > 0 \nonumber \\
 &\cdots& \nonumber \\
\partial^{(m)}\epsilon^\mu &=& O({1 \over r^{m + \alpha}}),
\, \alpha > 0  \ .
\label{ep-restr}
\eea
$\omega^\mu_{\ \nu}$ is an infinitesimal Lorentz parameter, 
$\eta_{\rho\mu}\omega^\mu_{\ \nu} + \eta_{\nu\mu}\omega^\mu_{\ \rho} = 0$, 
and $a^\mu$ is an infinitesimal translation of the coordinates.

Note that, under an infinitesimal diffeomorphism (\ref{ep-restr}),
\bea
\d_D g_{\mu\nu} &=& \epsilon^\rho g_{\mu\nu,\rho} + 
g_{\mu\rho}\epsilon^\rho_{,\nu} +g_{\rho\nu}\epsilon^\rho_{,\mu}  
\nonumber \\
&=& {\cal O}(r) \times  {\cal O}({1 \over r^2}) 
+ \left((\eta_{\mu\rho}+ {\cal O}({1 \over r}))(\omega^\rho_{\ \nu}+ {\cal O}
({1 \over r^2})) \right) 
\nonumber \\
&+& \left((\eta_{\rho\nu}+ {\cal O}({1 \over r}))(\omega^\rho_{\ \mu}+{\cal O}
({1 \over r^2})) \right) \nonumber \\
&=&{\cal O}({1 \over r}),
\eea
in agreement with the asymptotic behavior (\ref{asym}). Also 
$\d_D \partial^{(m)}g_{\mu\nu} = {\cal O}({1 \over r^{(m+1)}})$.

Let us check that (\ref{asym}) and (\ref{ep-restr}) 
imply the boundary conditions (\ref{boundcond}). 
The integration in (\ref{boundcond}) is, 
for ${\cal L}_K$,
\bea
B.C. &:=&\int_{\partial{}^3V} d\sigma_i \  
\left({\d S_K \over \d \phi_{i}}\d  \phi 
+ \partial_j({\d S_K \over \d \phi_{ij}} \d  \phi)\right) = 
\int_{\partial{}^3V} d\sigma_i  \ \left({\d S_K \over \d \phi_{i}}\d  \phi 
+ {1 \over 2}\partial_j({\d^c S_K \over \d \phi_{ij}} \d  \phi)\right) 
\nonumber \\
&=&\int_{\partial{}^3V} d\sigma_i  \ \left(P^{ab}(i)\d g_{ab}
- {1 \over 2}\partial_j(\alpha(ij) 
\sqrt{|\gamma(ij)|}\gamma^{AB}(ij) \d g_{AB} ) \right) \nonumber \\
&=& \int_{\partial{}^3V} d\sigma_i  \ \left(P^{ab}(i)\d g_{ab}
- \partial_j(\alpha(ij) 
 \d \sqrt{|\gamma(ij)|} ) \right) \ ,
\label{boundcondk}
\eea
where we have used (\ref{bulkmom}) and (\ref{boundmom}).

Since, when $r \rightarrow \infty $, the area of the boundary 
${\partial{}^3V}$ grows as ${\cal O}(r^2)$, the asymptotic 
behavior of (\ref{boundcondk}) is
\bea
\lim_{r \rightarrow \infty} B.C. &=&\lim_{r \rightarrow \infty}
\int_{\partial{}^3V} d\sigma_i  \ \left(P^{ab}(i)\d g_{ab}
- \partial_j(\alpha(ij)) 
 \d \sqrt{|\gamma(ij)|}  
- \alpha(ij) 
  \partial_j (\d\sqrt{|\gamma(ij)|})  \right) \nonumber \\
&=& \lim_{r \rightarrow \infty} {\cal O}(r^2) \left({\cal O}({1\over r^2})
\times{\cal O}({1\over r}) 
- {\cal O}({1\over r^2}) \times{\cal O}({1\over r}) 
- {\cal O}({1\over r})\times {\cal O}({1\over r^2})     
\right) \nonumber \\
&=& \lim_{r \rightarrow \infty} {\cal O}({1\over r}) = 0 \ ,
\label{boundcondk0}
\eea
in agreement with (\ref{boundcond}). It remains now to check that there is
consistency of the conditions (\ref{asym}) with the 
equations of motion (\ref{thedyn}) ---for finite times. 
This can be verified using the Lagrangian
equations of motion for ${\cal L}_{K}$ ---or ${\cal L}_{EH}$. The conditions 
of Proposition 1 are therefore satisfied.

\vspace{5mm}
Let us apply to ${\cal L}_{K}$ the results of the preceding section. 
The bulk momentum is obtained
from (\ref{bulkmom}),
\beq
{\d S_K \over \d  \dot g_{ij}} = P^{ij}(0) =: P^{ij} \ ,
\label{bulkmom0}
\eeq
for $i,j = 1,2,3$; and the boundary momentum form (\ref{boundmom}),
\beq
{\d S_K \over \d \dot g_{AB,i}} 
=  - \alpha(0i) \sqrt{|\gamma(0i)|}\gamma^{AB}(0i) \ ,
\label{boundmom0}
\eeq
for $A,B = 1,2,3 \ except \ i$. Let us find the Liouville 
and the presymplectic forms. The Liouville form is
\bea
\hat p_k d q^k &=& \int_{{}^3V} d^3\!x  \ ( {\d S_K \over \d \dot \phi}
\d  \phi) 
+ \int_{\partial{}^3V} d\sigma_i  \ 
( {\d S_K \over \d \dot \phi_{i}} \d \phi) \nonumber \\
&=& \int_{{}^3V} d^3\!x ( P^{ij}  \d g_{ij} )
- \int_{\partial{}^3V} d\sigma_i  \ 
( \alpha(0i) \sqrt{|\gamma(0i)|}\gamma^{AB}(0i) \d g_{AB}) 
\nonumber \\
&=& 
\int_{{}^3V} d^3\!x ( P^{ij} \d g_{ij} )
- 2 \int_{\partial{}^3V} d\sigma_i  \ 
( \alpha(0i) \d \sqrt{|\gamma(0i)|}) \ ,
\label{liouville}
\eea 
and the presymplectic form (\ref{presympl}) is, accordingly,
\bea
\hat \Omega &=& \int_{{}^3V} d^3\!x ( \d {\d S_K \over \d \dot \phi}
\wedge \d  \phi) 
+ \int_{\partial{}^3V} d\sigma_i  \ 
(\d {\d S_K \over \d \dot \phi_{i}}\wedge\d \phi) \nonumber \\
&=& 
\int_{{}^3V} d^3\!x  \ ( \d P^{ij} \wedge \d g_{ij} )
- 2 \int_{\partial{}^3V} d\sigma_i  \ 
(\d \alpha(0i)\wedge\d \sqrt{|\gamma(0i)|}) \ .
\label{presympl0}
\eea
This result was obtained in \cite{kijowski84,kijowski86} using the theory 
of symplectic relations \cite{tulc74, tulc78, kij79}; here it has been 
derived as a particular case of (\ref{presympl}).

According to the remarks produced at the end of 
subsection 4.2, since ${\cal L}_{K}$ differs from the 
scalar density Lagrangian ${\cal L}_{EH}$ by
boundary terms, 
it is already guaranteed that Noether currents exist for all diffeomorphism
symmetries. The point is that any Lagrangian ${\cal L}_{\rm any}$ differing
from ${\cal L}_{EH}$, a scalar density, by a divergence term,
$$
{\cal L}_{\rm any} = {\cal L}_{EH} + \partial_\mu D^\mu \ ,
$$
transforms, under an infinitesimal diffeomorphism $\d_D$ generated by the
vector field $\epsilon^\mu \partial_\mu$ (not necessarily satisfying 
(\ref{ep-restr})), as
$$
\d_D {\cal L}_{\rm any} = \partial_\mu(\epsilon^\mu {\cal L}_{EH}
+ \d_D D^\mu).
$$

\subsection{The energy}

Let us consider now the energy for a theory described by ${\cal L}_{K}$ plus
matter terms in an asymptotically flat spacetime. We will assume that 
these matter terms are described by a first order scalar density 
matter Lagrangian  
that together with ${\cal L}_{K}$ define the total Lagrangian,
and that the couplings with the metric are nonderivative. 
We will
also assume that the matter terms satisfy the appropriate conditions at the 
boundary in order to comply with the variational principle. 
We will first work with pure gravity and at the end consider the changes 
introduced by the presence of matter. So we continue with the 
Lagrangian ${\cal L}_K$.

Expression (\ref{energy}) gives, for the energy density ${\cal E}$ 
($E = \int_{{}^3V} d^3\!x \ {\cal E}$) ,
\bea
{\cal E} &=&   {\d S \over \d \dot \phi} \dot  \phi - {\cal L}
+ \partial_i ({\d S \over \d \dot \phi_{i}}\dot \phi)
\nonumber \\
&=& P^{ij} \dot g_{ij} - {\cal L}_{K}
-  2\partial_i ( \alpha(0i) \partial_0 \sqrt{|\gamma(0i)|}).
\label{energyk}
\eea
Introducing (\ref{tracelag}) and recalling $P^{ij} g_{ij} = 
2 \sqrt{|\gamma(0)|} K(0)$,
\bea
{\cal E} &=&  P^{ij} \dot g_{ij} - \sqrt{|g|} R
- 2 \partial_i(\sqrt{|\gamma(i)|} K(i)) - \partial_0( P^{ij} g_{ij}) 
\nonumber \\
&+& \partial_i \partial_j(\sqrt{|\gamma(ij)|} \alpha(ij)) 
+ 2 \partial_i \partial_0(\sqrt{|\gamma(0i)|} \alpha(0i))
-  2\partial_i ( \alpha(0i) \partial_0 \sqrt{|\gamma(0i)|}) \nonumber \\
&=& - \dot  P^{ij} g_{ij} - \sqrt{|g|} R 
-  2 \partial_i(\sqrt{|\gamma(i)|} K(i)) \nonumber \\
&+& \partial_i \partial_j(\sqrt{|\gamma(ij)|} \alpha(ij)) 
 + 2\partial_i ( \sqrt{|\gamma(0i)|} \partial_0 \alpha(0i))
\label{energydens}.
\eea
Let us now rewrite some terms in (\ref{energydens}).  
We show in the appendix, using the methods 
of \cite{kijowski97}, that
\beq
- \dot  P^{ij} g_{ij} + 
2\partial_i ( \sqrt{|\gamma(0i)|} \partial_0 \alpha(0i))
= 2\sqrt{|g|} R_0^{\, 0} - 2\partial_i( \sqrt{|g|}\gamma^{0\mu}(i) \,
\Gamma^i_{0\mu}) 
\label{kijformula}
\eeq
where $R_0^{\, 0}$ is a component of the Ricci tensor.
On the other hand, using (\ref{angles}), another piece in (\ref{energydens})
can be given a more convenient expression
\bea
\sqrt{|\gamma(ij)|} \partial_j \alpha(ij) &=& {1 \over 2}
\sqrt{|g|} \left( g^{ii} \partial_j ({g^{ij}\over g^{ii}}) + g^{jj} 
\partial_j ({g^{ij}\over g^{jj}})\right) \nonumber \\ 
&=& - \sqrt{|g|}(\Gamma^i_{j\mu}\gamma^{\mu j}(i) + 
\Gamma^j_{j\mu}\gamma^{\mu i}(j)) \ ,
\label{moreconvenient}
\eea 
where $\nabla_\mu g^{\rho\sigma} = 0$ has been used in the last equality.
Also, trivially,
\beq
\sqrt{|\gamma(i)|} K(i) = - \sqrt{|g|} \gamma^{\mu\nu}(i)\Gamma^i_{\mu\nu}.
\label{kagamma}
\eeq

All together,
\bea
{\cal E} &=& \sqrt{|g|}(2 R_0^{\, 0} - R) + 
2 \partial_i(\sqrt{|g|} \gamma^{\mu\nu}(i)\Gamma^i_{\mu\nu}) 
- \partial_i\left (\sqrt{|g|}(\Gamma^i_{j\mu}\gamma^{\mu j}(i) + 
\Gamma^j_{j\mu}\gamma^{\mu i}(j))\right) \nonumber \\ 
&+& \partial_i(\alpha(ij) \partial_j \sqrt{|\gamma(ij)|} )
- 2\partial_i( \sqrt{|g|}\gamma^{0\mu}(i)\Gamma^i_{0\mu}) \nonumber \\ 
&=& 2  G_0^{\, 0}
+ \partial_i(\alpha(ij) \partial_j \sqrt{|\gamma(ij)|} ) 
+ \partial_i\left (\sqrt{|g|}(\Gamma^i_{j\mu}\gamma^{\mu j}(i) - 
\Gamma^j_{j\mu}\gamma^{\mu i}(j))\right) \ .
\label{energydens2}
\eea

The first term in (\ref{energydens2}), a component of the Einstein tensor,  
vanishes on shell ---it is a
Lagrangian constraint, part of the equations of motion. 
The total energy on shell is
\bea
E &\sub{\rm (on \ shell)}&  \int_{{}^3V} d^3\!x \ 
{\cal E}\big|_{\rm on \ shell}
\nonumber \\
&=& \lim_{r \rightarrow \infty} \int_{\partial{}^3V} d\sigma_i \ 
\left(\alpha(ij) \partial_j \sqrt{|\gamma(ij)|} 
+ \sqrt{|g|}(\Gamma^i_{j\mu}\gamma^{\mu j}(i) - 
\Gamma^j_{j\mu}\gamma^{\mu i}(j))\right)  \ .
\label{asymenenrgy}
\eea

Considering the asymptotic behavior (\ref{asym}), the 
contribution of the first term in (\ref{asymenenrgy}) 
vanishes because
$\alpha(ij) = {\cal O}({1\over r})$. The second term contributes
\beq
E \sub{\rm (on \ shell)}  \
\lim_{r \rightarrow \infty}  
\int_{\partial{}^3V} d\sigma_i \ 
(\Gamma^i_{jj} - \Gamma^j_{ij})
= \
\lim_{r \rightarrow \infty}
\int_{\partial{}^3V} d\sigma_i \ (\partial_j g_{ij}
- \partial_i g_{jj}) \ ,
\label{adm}
\eeq
which is the ADM \cite{arnowitt/deser/misner/62} energy.  

The inclusion of matter, with the restrictions set at the beginning of this 
section, will change the term $G_0^{\, 0}$ to $G_0^{\, 0} - 8 \pi T_0^{\, 0}$,
where $T_0^{\, 0}$ is a component of the energy momentum tensor for 
the matter fields. Now $G_0^{\, 0} - 8 \pi T_0^{\, 0}$ is a constraint that
vanishes in virtue of Einstein equations. The asymptotic contribution 
in (\ref{adm}) remains the same.

\subsection{Other Noether charges for ${\cal L}_K$ }

Space translations in the $j$ direction (parameter $a^j$ in 
(\ref{ep-restr})) define for ${\cal L}_K$ the
quantity $F_{(j)}^i$ in (\ref{noethercase}),
$$
F_{(j)}^i = \d^i_j {\cal L}_K  a^j
$$
that should satisfy (\ref{condcons}) in Proposition 1. The  
$r \rightarrow \infty$ behavior of ${\cal L}_K$ can be easily deduced
from the expression (\ref{secorder}). The terms quadratic in the derivatives
of the metric behave as ${\cal O}({1 \over r^2}) \times 
{\cal O}({1 \over r^2}) =  {\cal O}({1 \over r^4})$. The term 
$\alpha(\mu\nu) \partial_\mu\partial_\nu(\sqrt{|\gamma(\mu\nu)|})$ has
$$
 \alpha(\mu\nu)  \subs{r \rightarrow \infty} {\cal O}({1 \over r}) \ ,
$$
and 
$$
\partial_\mu\partial_\nu(\sqrt{|\gamma(\mu\nu)|})
\subs{r \rightarrow \infty}  {\cal O}({1 \over r^3}) \ .
$$
Therefore
$$
{\cal L}_K \subs{r \rightarrow \infty} {\cal O}({1 \over r^4}) \ ,
$$
and so, 
\beq
\lim_{r \rightarrow \infty} \int_{\partial{}^3V} d\sigma_i \ {\cal L}_K = 0 \ .
\eeq
This proves that the translational Noether symmetry leads to conserved
quantities, the momenta. 

This result can be generalised: any 
diffeomorphism satisfying (\ref{ep-restr}) has a conserved charge. 
To prove it we need to compute $F^\mu$ in (\ref{noethercase}) for 
 an infinitesimal diffeomorphism $\d_D$ generated by 
$\epsilon^\mu \partial_\mu$.
This computation is involved because ${\cal L}_K$ is not a scalar density. 
The definition (\ref{tracelag}) obviously gives 
\beq
\d_D {\cal L}_{K} = \partial_\mu(\epsilon^\mu {\cal L}_{EH}) 
+ 2 \partial_\mu\left(\d_D(\sqrt{|\gamma(\mu)|} K(\mu))\right)
-  \partial_\mu \partial_\nu\left(\d_D(\sqrt{|\gamma(\mu\nu)|}  
\alpha(\mu\nu))\right) \ .
\label{dtracelag}
\eeq
but the behavior in the limit $r \rightarrow \infty$ of the first term 
$$
\epsilon^\mu {\cal L}_{EH} \subs{r \rightarrow \infty} 
{\cal O}({1 \over r^2}) \ ,
$$
(with $\epsilon^\mu \subs{r \rightarrow \infty} {\cal O}(r)$ ) 
produces terms different from zero in (\ref{condconsinf}). In fact this
behavior is corrected by other terms in (\ref{dtracelag}) but to single out
these contributions is cumbersome. It is much more convenient to express 
${\cal L}_K$ as the sum of the ``gamma-gamma'' Lagrangian ${\cal L}_{\Gamma}$
plus divergences, because the behavior of ${\cal L}_{\Gamma}$ 
in the limit $r \rightarrow \infty$ is much better than that of 
${\cal L}_{EH}$,
$$
\epsilon^\mu {\cal L}_{\Gamma}  \subs{r \rightarrow \infty}
{\cal O}({1 \over r^3}) \ . 
$$

So let us proceed to relate ${\cal L}_K$ with ${\cal L}_{\Gamma}$. 
Using (\ref{kagamma}), a trivial extension of (\ref{moreconvenient}) 
to indices $\mu\nu$,
\bea
\partial_\mu\left(
\sqrt{|\gamma(\mu\nu)|} \partial_\nu \alpha(\mu\nu)
\right) &=& - \partial_\mu\left(\sqrt{|g|}
(\Gamma^\mu_{\nu\sigma}\gamma^{\nu\sigma}(\mu)  
+ \Gamma^\nu_{\nu\sigma}\gamma^{\sigma\mu}(\nu))\right) \nonumber \\
&=& 2 \partial_\mu\left(\sqrt{|\gamma(\mu)|}K(\mu)\right)
+ \partial_\mu\left(\sqrt{|g|}(\Gamma^\mu_{\nu\sigma}\gamma^{\nu\sigma}(\mu) - 
\Gamma^\nu_{\nu\sigma}\gamma^{\sigma\mu}(\nu))\right) \ ,
\eea 
allows to write ${\cal L}_K$ in the equivalent form
\beq
{\cal L}_K = \sqrt{|g|}R - \partial_\mu\left(\alpha(\mu\nu)
\partial_\nu \sqrt{|\gamma(\mu\nu)|} 
\right)  
-\partial_\mu\left(\sqrt{|g|}(\Gamma^\mu_{\nu\sigma}\gamma^{\nu\sigma}(\mu) - 
\Gamma^\nu_{\nu\sigma}\gamma^{\sigma\mu}(\nu))\right) \ . 
\eeq
On the other hand, the ``gamma-gamma'' Lagrangian (\ref{gammaaction})
can be written as 
\beq 
{\cal L}_{\Gamma} = \sqrt{|g|}R 
-\partial_\mu\left(\sqrt{|g|}(\Gamma^\mu_{\nu\sigma}g^{\nu\sigma} - 
\Gamma^\nu_{\nu\sigma}g^{\sigma\mu})\right) \ ,
\label{gammagamma} 
\eeq
and therefore, using (\ref{gammag}), 
\bea
{\cal L}_K &=& {\cal L}_{\Gamma} - \partial_\mu\left(\alpha(\mu\nu)
\partial_\nu \sqrt{|\gamma(\mu\nu)|} 
\right) \nonumber \\ 
&+&\partial_\mu\left(\sqrt{|g|}(
 \eta(\mu) n^\sigma(\mu) n^\nu(\mu)\Gamma^\mu_{\nu\sigma} -  
 \eta(\nu) n^\sigma(\nu) n^\mu(\nu)\Gamma^\nu_{\nu\sigma}
)\right) \ . 
\eea

We continue with $\d_D$ being an infinitesimal diffeomorphism generated by 
$\epsilon^\mu \partial_\mu$. Computing $\d_D{\cal L}_{\Gamma}$ is a simple 
thing \cite{faddeev82} because the deviation of ${\cal L}_{\Gamma}$ from the 
scalar density behavior is due exclusively to the two connexion 
coefficients present in the divergence 
term in (\ref{gammagamma}). Since the deviation of the connexion 
$\Gamma^\mu_{\nu\sigma}$ from a tensorial behavior is an additive term
$\epsilon^\mu_{,\nu\sigma}$, we can write 
$$
\d_D {\cal L}_{\Gamma} = \partial_\mu\left( \epsilon^\mu{\cal L}_{\Gamma}
- \sqrt{|g|}g^{\nu\sigma}\epsilon^\mu_{,\nu\sigma}
+  \sqrt{|g|}g^{\mu\sigma}\epsilon^\nu_{,\nu\sigma}\right) \ ,
$$
and thus, for ${\cal L}_K$,
\bea
\d_D{\cal L}_K &=&\partial_\mu\Biggl( \epsilon^\mu{\cal L}_{\Gamma}
- \sqrt{|g|}g^{\nu\sigma}\epsilon^\mu_{,\nu\sigma}
+  \sqrt{|g|}g^{\mu\sigma}\epsilon^\nu_{,\nu\sigma} 
- \d_D \left(\alpha(\mu\nu)
\partial_\nu \sqrt{|\gamma(\mu\nu)|} 
\right) \nonumber \\ 
&+& \d_D\left(\sqrt{|g|}(
 \eta(\mu) n^\sigma(\mu) n^\nu(\mu)\Gamma^\mu_{\nu\sigma} -  
 \eta(\nu) n^\sigma(\nu) n^\mu(\nu)\Gamma^\nu_{\nu\sigma}
)\right)\Biggr) \ . 
\label{deltalk}
\eea

Now, from (\ref{deltalk}), we deduce the object $F^i$ that needs to pass the
test (\ref{condconsinf}),
\bea
F^i &=& \epsilon^i{\cal L}_{\Gamma}
- \sqrt{|g|}g^{\nu\sigma}\epsilon^i_{,\nu\sigma}
+  \sqrt{|g|}g^{i\sigma}\epsilon^\nu_{,\nu\sigma} 
- \d_D \left(\alpha(i\nu)
\partial_\nu \sqrt{|\gamma(i\nu)|} 
\right) \nonumber \\ 
&+& \d_D\left(\sqrt{|g|}(
 \eta(i) n^\sigma(i) n^\nu(i)\Gamma^i_{\nu\sigma} -  
 \eta(\nu) n^\sigma(\nu) n^i(\nu)\Gamma^\nu_{\nu\sigma}
)\right) \ .
\label{efforlk}
\eea
Now it is easy to check that each additive term in (\ref{efforlk}), under
the conditions (\ref{asym}) and  (\ref{ep-restr}),
guarantees (\ref{condconsinf}) and therefore
$$
\lim_{r \rightarrow \infty} 
\int_{\partial{}^3V} d\sigma_i \  F^i  = 0 \ ,
$$ 
thus proving that all diffeomorphisms satisfying (\ref{ep-restr}) give
Noether conserved charges for asymptotically flat spaces defined by  
the conditions (\ref{asym}). Note in particular the existence of
conserved charges associated with the Poincar\'e transformations
included in (\ref{ep-restr}). In this sense, (\ref{adm}) is just an example 
of the computation ---the total energy--- of one of the ten conserved charges
corresponding to the Poincar\'e group.  
\section{Conclusions}

In this paper we have developed the formal theory for the 
variational principle and the Noether symmetries for second order 
Lagrangians in field theories with boundaries, with special emphasis
in Lagrangians with no second order time derivatives. These developments 
lead to general expressions for some geometric objects, like the pullback
to tangent space of the symplectic form in  phase space. It is worth
to remark that bulk terms and boundary terms contribute to these objects. Also
the canonical momenta exhibit both a bulk piece and a boundary piece.

The Noether theorem is discussed for these theories. A clear distinction
is made between conservation of currents, that is a local assertion only
related to the equations of motion ---but not to the possible boundary terms 
in the action---, and the conservation of charges, where the relation with
the boundary terms appearing in the action is made transparent. The reason 
being that the boundary terms in the action are linked to the boundary 
conditions to be satisfied by the fields. These results are summarised 
in a proposition introduced in subsection 4.2. Let us mention also that
the Noether charge exhibits a bulk piece and a 
boundary piece, the boundary piece owing its existence to the second order 
dependences in the Lagrangian.

This framework is applied to the trace K Lagrangian for General Relativity.
The trace K Lagrangian is obtained in section 3 and its applications
to asymptotically flat spaces are studied in section 5. We observe that 
this Lagrangian is not first order in the derivatives but its asymptotic
behavior is different from that of the Einstein-Hilbert 
Lagrangian and resembles that of first order Lagrangians like the 
gamma-gamma Lagrangian. We prove the 
conservation of charges for diffeomorphisms that preserve the boundary 
conditions for the metric tensor and, as a particular case, we obtain the ADM
formula for the conserved energy.

It is worth noting that we have based our approach in keeping at
any moment the requirements derived from the strict application of the 
variational principle, including
in particular the differentiability of the action functional. These
requirements lead to the imposition of boundary conditions on the field 
configurations. When we examine the conditions for the conservation of
possible Noether charges the relevance of these boundary conditions, and
hence of the variational principle, becomes transparent.    

\vspace{8mm}

Let us finally give a short review of the main results 
obtained in this paper.

\begin{enumerate}

\item

\underline{General Theory}

\begin{itemize}

\item
In Proposition 1, section 4.2, we have given the necessary and sufficient 
conditions for the conservation of Noether 
charges for field theories with boundaries; these field theories being 
derived from 
action principles with second order Lagrangians that are free from  
second order time derivative terms.  

\item
We have shown that the Noether conserved charges, (\ref{truecharge}), 
exhibit a boundary piece that stems from the presence of the 
one-time-one-space derivative terms in the Lagran\-gian. 

\item
We have derived a general formula, (\ref{presympl}), for the 
(pre)sym\-plec\-tic
form associated with this type of Lagrangians. This form exhibits a boundary 
piece that has the same origin as for the Noether charges. 

\item
Also, due to the presence of the one-time-one-space 
derivative terms in the Lagrangian, we show that one is naturally led to 
define a bulk component, (\ref{bulkm}), as well as a  boundary 
component, (\ref{boundm}), for each of the momenta.

\end{itemize}

\item
\underline{Trace K Lagrangian}

\begin{itemize}
\item
We have neatly displayed, (\ref{secorder}), the dependence of the Trace K 
Lagrangian on second order spacetime derivatives. We also show that the 
asymptotic behavior of this Lagrangian is similar to the one of the 
first order ``gamma-gamma'' Lagrangian, substantially different from that of 
the Einstein-Hilbert Lagrangian.

\item
We have applied our general formulas to compute the (pre)symplectic form 
associated with the Trace K Lagrangian, (\ref{presympl0}). 
Our results coincide with 
those obtained within the theory of symplectic relations.

\item
We have given an expression for the energy density, (\ref{energydens2}), 
and we have computed the total energy for the asymptotically flat case, 
obtaining the ADM formula.

\item
We have shown that, in the asymptotically flat case, all diffeomorphisms 
satisfying the boundary conditions -this includes the transformations 
that are asymptotically Poin\-car\'e- give Noether conserved charges. 
      
\end{itemize}

\end{enumerate}
\section{Appendix}
\subsection{Deviation of the transformation 
of ${\bf n}(\mu)$ from the vector behavior}

Let $\d_D$ be the infinitesimal diffeomorphism generated by 
$\epsilon^\mu \partial_\mu$. Our task is to compute $\d_D n^\nu(\mu)$ for
the ``vector'' ${\bf n}(\mu) $whose components, $n^\nu(\mu)$, are 
defined through  
$$
n^\mu(\mu) n^\nu(\mu) = \xi(\mu) g^{\mu\nu} \ ,
$$ 
and we know
$$
\d_D g^{\mu\nu} = \epsilon^\sigma \partial_\sigma g^{\mu\nu}
- g^{\mu\sigma}\partial_\sigma\epsilon^\nu 
- g^{\sigma\nu}\partial_\sigma\epsilon^\mu \ . 
$$

Considering 
$$
(n^\mu(\mu))^2 = \xi(\mu) g^{\mu\mu} \ ,
$$
we obtain
$$
\d_D n^\mu(\mu) = \d_{\rm naive} n^\mu(\mu) \ ,
$$
where we have introduced $\d_{\rm naive}$ to symbolise a ``naive'' 
variation associated with vector behavior, that is, 
$$
\d_{\rm naive} n^\nu(\mu) := \epsilon^\sigma \partial_\sigma n^\nu(\mu)
- n^\sigma (\mu)\partial_\sigma\epsilon^\nu.
$$

Next, to find  $\d_D n^\nu(\mu)$, use
$$
n^\nu(\mu) = \xi(\mu) g^{\mu\nu} {1 \over n^\mu(\mu)} \ ,  
$$
and so,
$$
\d_D n^\nu(\mu)= \xi(\mu)\d_D g^{\mu\nu} {1 \over n^\mu(\mu)} 
+ \xi(\mu)g^{\mu\nu} \d_D {1 \over n^\mu(\mu)} \ . 
$$
This expression can be arranged to give
$$
\d_D n^\nu(\mu) = \d_{\rm naive} n^\nu(\mu) 
- \Bigr(n_\mu(\mu) g^{\nu\sigma} 
- \xi(\mu) n_\mu(\mu)n^\nu(\mu) n^\sigma(\mu) \Bigl)
\partial_\sigma\epsilon^\mu \ , 
$$
which, using (\ref{gammag}), is
\bea
\d_D n^\nu(\mu) &=& \d_{\rm naive} n^\nu(\mu) 
- n_\mu(\mu) \gamma^{\nu\sigma}(\mu) \partial_\sigma\epsilon^\mu \nonumber \\
&=& \d_{\rm naive} n^\nu(\mu) 
- n_\mu(\mu) \d_a^\nu\gamma^{ab}(\mu) \partial_b\epsilon^\mu \ ,
\label{deviation}
\eea
for $a,b = 0,1,2,3 \ except \  \mu$. 
Equation (\ref{deviation}) expresses the deviation of the transformation 
of ${\bf n}(\mu)$ from the ``naive'' vector behavior. Notice that this 
deviation differs from zero only for infinitesimal diffeomorphisms 
$\epsilon^\mu$ such that  
$\partial_b\epsilon^\mu \neq 0$; these are the diffeomorphisms that
do not preserve the foliation of ${}^4\!V$ 
in surfaces $x^\mu = constant$.

\subsection{Proof of (\ref{secorder})}

Let us single out, for instance, the $\partial_0\partial_3$ terms in 
$$
{\cal L}_{K} = \sqrt{|g|}R + 2 \partial_\mu(\sqrt{|\gamma(\mu)|} K(\mu))
-  \partial_\mu \partial_\nu(\sqrt{|\gamma(\mu\nu)|}  \alpha(\mu\nu)) \ .
$$
To this end, it is convenient to use the ADM decomposition
$$
\sqrt{|g|}R = {\cal L}_{ADM} - 2\partial_\mu\left(\sqrt{|g|}
(n^\lambda(0) n^\mu_{;\lambda}(0) - n^\mu(0) n^\lambda_{;\lambda}(0)) \right)
$$
where 
$$
{\cal L}_{ADM} := -\sqrt{|\gamma(0)|} n_0(0)({}^3\!R + K_{ab}(0)K^{ab}(0) \,
- K^2(0))
$$
is free from $\partial_0\partial_3$ terms. ${}^3\!R$ is the scalar curvature
for the surface $x^0 = constant$. 

We will use the notation $[something]_{|_\mu}$ to isolate the additive terms
in $[something]$ that contain a $\partial_\mu$ derivative, and similarly for
$[something]_{|_{\mu\nu}}$. So
\bea
[{\cal L}_{K}]_{|_{03}} &=& - 2\partial_0\left(\sqrt{|g|}
[n^\lambda(0) n^0_{;\lambda}(0) 
- n^0(0) n^\lambda_{;\lambda}(0)]_{|_3} \right) \nonumber \\
&-& 2\partial_3\left(\sqrt{|g|}
[n^\lambda(0) n^3_{;\lambda}(0) - 
n^3(0) n^\lambda_{;\lambda}(0)]_{|_0} \right) \nonumber \\
&+& 2 \partial_0\left([\sqrt{|\gamma(0)|} K(0)]_{|_3}\right)
+ 2 \partial_3\left([\sqrt{|\gamma(3)|} K(3)]_{|_0}\right)
- 2 \partial_0 \partial_3(\sqrt{|\gamma(03)|}  \alpha(03)) \ .
\eea

The first and third term in the right side cancel because $n^0_{;\lambda}(0)=0$
and $n^\lambda_{;\lambda}(0) = - K(0)$. On the other hand, the piece in the
second term $n^\lambda(0) n^3_{;\lambda}(0)$ can be expressed as 
$n^\lambda(0) n^3_{;\lambda}(0) = n_0(0) \gamma^{3j} \partial_j n^0(0)$, 
that has no $\partial_0$ derivative. 
Therefore
\beq
[{\cal L}_{K}]_{|_{03}} =
 2\partial_3\left([ 
n^3(0) n^\lambda_{;\lambda}(0) + 
  \sqrt{|\gamma(3)|} K(3) 
-  \partial_0(\sqrt{|\gamma(03)|}  \alpha(03))]_{|_0}\right) \ ,
\label{zerothree}
\eeq

Now, a little algebra gives
\bea
[n^3(0) n^\lambda_{;\lambda}(0) + 
  \sqrt{|\gamma(3)|} K(3)]_{|_0} &=&  \sqrt{|\gamma(3)|} n^0(3)
(ln{n^0(3) \over n^0(0)})_{,0} \nonumber \\
&=&  \sqrt{|\gamma(03)|} { 1 \over 
\sqrt{1 + q^2(03)}} \partial_0 q(03) \nonumber \\
&=& \sqrt{|\gamma(03)|}\partial_0 \alpha(03) \ ,
\eea
and, finally, plugging this result in (\ref{zerothree}), 
\beq
[{\cal L}_{K}]_{|_{03}} =
 -  2 \alpha(03) \partial_0\partial_3(\sqrt{|\gamma(03)|}) \ . 
\eeq

The ADM decomposition gives also an immediate proof that  
$\partial_0\partial_0$ terms are not present in ${\cal L}_{K}$.
An extension \cite{york86} of the ADM 
decomposition ---which is associated with the $\mu = 0$ coordinate--- 
to any other coordinate helps to prove, along the same lines, that  
\beq
[{\cal L}_{K}]_{|_{\mu\mu}} = 0 \ , 
\eeq
and that 
\beq
[{\cal L}_{K}]_{(\rm {second \ order \ terms})} =
 -  \alpha(\mu\nu) \partial_\mu\partial_\nu(\sqrt{|\gamma(\mu\nu)|}) \ .
\eeq

\subsection{Proof of (\ref{kijformula})}
Recalling (\ref{kijfirst}), for $\d$ being the time derivative, 
\beq
\tilde g^{\mu\nu 0}_\sigma \dot \Gamma^\sigma_{\mu\nu} =
- g_{ij} \dot P^{ij} + \partial_i (\sqrt{|g|} g^{00} 
\partial_0 ({g^{0i}\over g^{00}})) \ ,
\eeq
and recalling also (\ref{angles}) (with $\d$ being the time derivative)
\beq
\sqrt{|g|} \left( g^{00} 
\partial_0 ({g^{0i}\over g^{00}}) + g^{ii} 
\partial_0 ({g^{0i}\over g^{ii}})\right) =
2 \sqrt{|\gamma(0i)|} \partial_0 \alpha(0i) \ ,
\eeq
we can write
\beq
g_{ij} \dot P^{ij} - 2 \partial_i(\sqrt{|\gamma(0i)|} 
\partial_0 \alpha(0i))  
= - \tilde g^{\mu\nu 0}_\sigma \dot \Gamma^\sigma_{\mu\nu}
- \partial_i(g^{ii} 
\partial_0 ({g^{0i}\over g^{ii}}))\ .
\eeq

Considering \cite{kijowski97} the following identity with $X^\mu = (1,0,0,0)$,
that originates from the definition of the Riemann tensor,
\beq
\dot \Gamma^\sigma_{\mu\nu} = \nabla_\mu\nabla_\nu X^\sigma  
- R_{\lambda\mu\nu}^{\quad \ \sigma} X^\lambda \ ,
\eeq
then
\bea
\tilde g^{\mu\nu\rho}_\sigma \dot \Gamma^\sigma_{\mu\nu} &=&
\sqrt{|g|} (\nabla_\mu\nabla^\mu X^\rho - 
\nabla_\mu\nabla^\rho X^\mu - g^{\mu\nu} R_{\lambda\mu\nu}^{\quad \ \rho}
X^\lambda 
+ g^{\rho\nu} R_{\lambda\mu\nu}^{\quad \ \mu} X^\lambda ) \nonumber \\
&=& \partial_\mu \left(\sqrt{|g|}(\nabla^\mu X^\rho - \nabla^\rho X^\mu)\right)
+ 2 \sqrt{|g|} R_\lambda^{\ \rho} X^\lambda \ , 
\eea
and, for $X^\mu = (1,0,0,0)$ and $\rho = 0$,
\bea
\tilde g^{\mu\nu 0}_\sigma \dot \Gamma^\sigma_{\mu\nu}
&=&\partial_\mu \left(\sqrt{|g|}(\nabla^\mu X^0 - \nabla^0 X^\mu)\right)
+ 2 \sqrt{|g|} R_0^{\ 0}  \nonumber \\
&=&\partial_i \left(\sqrt{|g|}(\nabla^i X^0 - \nabla^0 X^i)\right)
+ 2 \sqrt{|g|} R_0^{\ 0}. 
\eea
Then, 
\bea
g_{ij} \dot P^{ij} - 2 \partial_i(\sqrt{|\gamma(0i)|} 
\partial_0 \alpha(0i))  &=&
 - \tilde g^{\mu\nu 0}_\sigma \dot \Gamma^\sigma_{\mu\nu}
- \partial_i(g^{ii} \partial_0 ({g^{0i}\over g^{ii}})) \nonumber \\
&=&
- 2 \sqrt{|g|} R_0^{\ 0} - \partial_i\left(\sqrt{|g|}
((\nabla^i X^0 - \nabla^0 X^i) 
+ g^{ii} \partial_0 ({g^{0i}\over g^{ii}}))\right) \nonumber \\
&=&
- 2 \sqrt{|g|} R_0^{\ 0} + 2 \partial_i\left(
\sqrt{|g|}\gamma^{0\mu}(i)\Gamma^i_{0\mu}\right).
\eea




\begin{thebibliography}{99}

\bibitem{Regge:1974zd}
T.~Regge and C.~Teitelboim,
Annals Phys.\  {\bf 88} (1974) 286.


\bibitem{Katz:1997si}
J.~Katz,
{\it 8th Marcel Grossmann Meeting on Recent Developments in Theoretical and Experimental General Relativity, Gravitation and Relativistic Field
Theories (MG 8), Jerusalem, Israel, 22-27 Jun 1997}.

\bibitem{Julia:1998ys}
B.~Julia and S.~Silva,
Class.\ Quant.\ Grav.\  {\bf 15} (1998) 2173
[arXiv:gr-qc/9804029].

\bibitem{Julia:2000er}
B.~Julia and S.~Silva,
Class.\ Quant.\ Grav.\  {\bf 17} (2000) 4733
[arXiv:gr-qc/0005127].

\bibitem{Brown:1993br}
J.~D.~Brown and J.~W.~York,
Phys.\ Rev.\ D {\bf 47} (1993) 1407.

\bibitem{Chen:1999aw}
C.~M.~Chen and J.~M.~Nester,
Class.\ Quant.\ Grav.\  {\bf 16} (1999) 1279
[arXiv:gr-qc/9809020].


\bibitem{Chang:1999wj}
C.~C.~Chang, J.~M.~Nester and C.~M.~Chen,
Phys.\ Rev.\ Lett.\  {\bf 83} (1999) 1897
[arXiv:gr-qc/9809040].


\bibitem{brown00} J.~D.~Brown, S.~R.~Lau and J.~W.~York,
[gr-qc/0010024]

\bibitem{dirac58} P. A. M. Dirac, 
{\sl Proc. Roy. Soc. (London) \bf{A 246}}, 333 (1958)

\bibitem{gibb77} G: W: Gibbons and S. W. Hawking,
{\sl Phys. Rev. \bf{D 15}}, 2752 (1977)

\bibitem{york72} J. W. York, 
{\sl Phys. Rev. Lett. \bf{28}}, 1082 (1972)


\bibitem{Charap:1983kn}
J.~M.~Charap and J.~E.~Nelson,
J.\ Phys.\ A {\bf 16} (1983) 1661.

\bibitem{york86} J. W. York,
{\sl Found. of Physics \bf{16}}, 249 (1986)  

\bibitem{gold87} J. N. Goldberg,
{\sl Phys. Rev {\bf D 37}}, 2116 (1988)

\bibitem{hayw93} G. Hayward,
{\sl Phys. Rev {\bf D 47}}, 3275 (1993)

\bibitem{Hawk96fd}
S.~W.~Hawking and G.~T.~Horowitz,
{\sl Class.\ Quant.\ Grav.\ {\bf 13}}, 1487 (1996)
[gr-qc/9501014]

\bibitem{hawk96} S.~W.~Hawking and C.~J.~Hunter,
{\sl Class.\ Quant.\ Grav.\ {\bf 13}}, 2735 (1996)
[gr-qc/9603050]

\bibitem{epp98} R.~J.~Epp and R.~B.~Mann,
{\sl Mod.\ Phys.\ Lett.\ A {\bf 13}}, 1875 (1998)
[gr-qc/9806004]


\bibitem{Franca02a}
L.~Fatibene, M.~Ferraris, M.~Francaviglia and M.~Raiteri,
J.\ Math.\ Phys.\  {\bf 42} (2001) 1173
[arXiv:gr-qc/0003019].


\bibitem{Franca02b}
M.~Francaviglia and M.~Raiteri,
Class.\ Quant.\ Grav.\  {\bf 19} (2002) 237
[arXiv:gr-qc/0107074].


\bibitem{kijowski97} J. Kijowski,
{\sl Gen. Rel. and Grav. \bf{29}}, 307 (1997)

\bibitem{kijowski84} J. Kijowski,
In {\sl Proc. Journ\'ees Relativistes (Torino 1983)} S. Benenti et al., eds. 
(Pitagora Editrice, Bologna), p. 205; (1984)

\bibitem{kijowski86} J. Kijowski,
In {\sl Gravitation, Geometry and Relativistic Physics} (Springer
Lecture Notes in Physics 212, Springer-Verlag, Berlin (1986)

\bibitem{arnowitt/deser/misner/62}
R.\ Arnowitt, S.\ Deser, and C.\ W.\ Misner,
in {\it Gravitation: An Introduction to Current Research}. 
Edited by L.\ Witten 
(John Wiley \& Sons, New York, 1962), 227--265 (1962)

\bibitem{brownhenn86} J.~D.~Brown and M. Henneaux,
{\sl J. Math. Phys. \bf 27}, 489 (1986) 

\bibitem{faddeev82}
L.~D.~Faddeev,
{\sl Sov.\ Phys.\ Usp. {\bf 25}}, 130 (1982). 

\bibitem{brill92} D. ~Brill,
{\sl Phys.Rev. {\bf D 46}}, 1560 (1992)


\bibitem{Marolf:1995cp}
D.~Marolf,
Phys.\ Lett.\ B {\bf 349} (1995) 89
[arXiv:gr-qc/9411067].

\bibitem{soloviev} V. O. Soloviev,
{\sl J. Math. Phys. \bf 34}, 5747 (1993) 

\bibitem{Pons:2000xt}
J.~M.~Pons, D.~C.~Salisbury and L.~C.~Shepley,
{\sl Phys.\ Rev.\ D {\bf 62}}, 064026 (2000)
[gr-qc/9912085]

\bibitem{sheikh99}
M.~M.~Sheikh-Jabbari and A.~Shirzad,
{\sl Eur.\ Phys.\ J.\ C {\bf 19}} (2001) 383
[hep-th/9907055]

\bibitem{tulc74} W. M. Tulczyjev,
{\sl Symposia mathematica \bf 14}, 247 (1974)

\bibitem{tulc78} M. R. Menzio and W. M. Tulczyjev,
{\sl Ann. Inst. H. Poincar\'e \bf 28}, 349 (1978)

\bibitem{kij79} J. Kijowski and W. M. Tulczyjev,
{\sl A Symplectic Framework for field Theories} 
(Lecture Notes in Physics 107,
Springer-Verlag, Berlin.) (1979)


\end{thebibliography}
\end{document}